\documentclass[10pt,letterpaper]{article}
\usepackage[top=0.85in,left=2.75in,footskip=0.75in]{geometry}

\usepackage{amsmath,amssymb}

\usepackage{changepage}

\usepackage{textcomp,marvosym}

\usepackage{cite}

\usepackage{nameref,hyperref}

\usepackage[nopatch=eqnum]{microtype}
\DisableLigatures[f]{encoding = *, family = * }

\usepackage[table]{xcolor}

\usepackage{array}

\usepackage{booktabs}

\newcolumntype{+}{!{\vrule width 2pt}}

\newlength\savedwidth

\raggedright
\usepackage[aboveskip=1pt,labelfont=bf,labelsep=period,justification=raggedright,singlelinecheck=off]{caption}

\makeatletter
\renewcommand{\@biblabel}[1]{\quad#1.}
\makeatother

\usepackage{lastpage,fancyhdr,graphicx}
\usepackage{epstopdf}
\fancyheadoffset[L]{2.25in}
\fancyfootoffset[L]{2.25in}
\begin{document}
\vspace*{0.2in}

\begin{flushleft}
{\Large
\textbf{Evaluating Small Open LLMs for Medical Question Answering: A Practical Framework}
}
\newline
\\
Avi-ad Avraam Buskila\textsuperscript{1,*}
\\
\bigskip
\textbf{1} Department of Information Science and Applied Artificial Intelligence, Bar-Ilan University, Ramat-Gan, Israel
\\
\bigskip

% Use the asterisk to denote corresponding authorship and provide email address in note below.
* aviad-avraam.buskila@biu.ac.il

\end{flushleft}
\section*{Abstract}
Deploying large language models (LLMs) for medical question answering
demands more than high average accuracy: a model that returns
substantively different answers each time it is queried is not a reliable
clinical tool, regardless of its mean quality score. In settings like online
health communities, where misinformation spreads
readily~\cite{suarezlledo2021misinfo, sager2021reddit}, output consistency is
as important as correctness. We present a framework for
evaluating small, locally-deployable open-weight LLMs on medical question
answering that treats \emph{reproducibility} as a first-class metric.
The pipeline computes quality metrics (BERTScore, ROUGE-L, and an
LLM-as-judge rubric) and complementary lexical and semantic within-model
reproducibility metrics over repeated inference ($N{=}10$ runs per question).
We evaluate four open-weight models (Llama 3.1 8B, Gemma 3 12B, Gemma 3 4B,
and the clinically fine-tuned MedGemma 1.5 4B), with the closed GPT-4.1-nano
as an external reference, on two consumer-health benchmarks (MedQuAD and
MedicationQA; 50 questions each, 10 runs per question, 5{,}000 responses in
total). Low-temperature generation ($T{=}0.2$) does not yield reproducible
outputs: lexical self-agreement is only 0.11--0.26 and 76--99\% of outputs
per model are unique. Yet the mean pairwise \emph{semantic} self-similarity
across runs is 0.94--0.97 for every model on both datasets, showing that this
variability is overwhelmingly paraphrastic rather than meaning-level---a
distinction purely lexical, single-pass benchmarks miss, and one that holds
even for the closed reference. With the same-size control included, model
scale rather than clinical fine-tuning tracks quality: MedGemma 1.5 4B is
statistically indistinguishable from its same-size sibling Gemma 3 4B, while
the larger Gemma 3 12B leads the open models. We report clustered bootstrap
CIs throughout and quantify the LLM judge's reliability
(ICC(1)~$=$~0.94 on MedQuAD, 0.69 on MedicationQA).

\clearpage
\newgeometry{top=0.85in,left=1in,right=1in,footskip=0.75in}

\section*{Introduction}

Online medical communities such as Reddit have become a primary
source of health information for millions of users, yet they remain
highly susceptible to the rapid spread of misinformation and unverified
advice~\cite{suarezlledo2021misinfo, sager2021reddit}.
While large language models (LLMs) offer a promising opportunity to support
question answering in these settings, their integration into
community-driven platforms requires careful evaluation of reliability
and safety. In this work, we explore the potential of small,
locally deployable open LLMs as assistants in medical subreddits,
with the goal of improving answer quality while mitigating the
risks associated with inconsistent or misleading model outputs.

The availability of capable open-weight LLMs that can run entirely
on local hardware has made clinical AI accessible to resource-constrained
settings where patient data cannot leave institutional boundaries.
Models such as Llama 3.1~\cite{dubey2024llama3},
Gemma 3~\cite{team2024gemma}, and domain-fine-tuned variants like
MedGemma~\cite{medgemma2024} can now be queried on a commodity
workstation, lowering both the cost and the compliance burden of
piloting LLM-based clinical decision support.

Yet the dominant evaluation paradigm for these models, benchmarking
single-pass accuracy against a clinically validated ground-truth corpus,
is fundamentally inadequate for medical usage.
A drug-interaction query that returns a correct answer 70\% of the time
and a plausible but wrong answer the other 30\% is not a useful medical
tool; neither is one that returns a different correct answer on every
invocation, because the instability itself erodes trust and
auditability.
In safety-critical applications, regulators and risk managers need to
know not only \emph{whether} a model is usually correct, but
\emph{how stable} those answers are.

Existing clinical NLP benchmarks, MedQA~\cite{jin2021disease},
PubMedQA~\cite{jin2019pubmedqa}, and others, evaluate single-pass
accuracy and leave run-to-run variance unmeasured.

This paper makes the following contributions:
\begin{enumerate}
  \item A \textbf{reproducibility-first evaluation protocol} that
        quantifies within-model run-to-run consistency via lexical
        self-agreement and response-uniqueness rates \emph{and} a
        semantic self-similarity measure, alongside a standard accuracy
        suite, with all estimates reported as bootstrap confidence
        intervals.
  \item A \textbf{modular, resumable pipeline} implementing the protocol
        in three stages (run\;$\to$\;score\;$\to$\;report).
  \item \textbf{Empirical results} on five models---four small
        open-weight models spanning 4B--12B parameters plus a closed
        reference---across two consumer-health datasets ($N{=}5{,}000$
        responses; 50 questions $\times$ 10 runs per model per dataset),
        demonstrating that reproducibility failures persist even in the
        near-deterministic temperature regime, that run-to-run
        divergence is largely paraphrastic (high semantic
        self-similarity despite low lexical agreement), and that---using
        a same-size control---model scale rather than clinical
        fine-tuning drives quality.
\end{enumerate}

% ---- 2. Related Work -----------------------------------------
\section*{Related work}
\label{sec:related}

\paragraph{Self-consistency and output stability.}
Wang et~al.~\cite{wang2022selfconsistency} introduced self-consistency
as a decoding strategy that samples multiple chain-of-thought paths
and selects the most frequent answer, substantially improving reasoning
accuracy. Chen et~al.~\cite{chen2023universal} extended this idea to
free-form generation with universal self-consistency.
Both lines of work use repeated sampling to \emph{improve} output quality;
our work repurposes the same repeated-sampling setup to \emph{measure}
output stability, turning the consistency signal itself into a
first-class evaluation metric rather than a decoding technique.

\paragraph{LLM calibration and reliability.}
Kadavath et~al.~\cite{kadavath2022calibration} showed that language
models can partially assess the correctness of their own outputs, yet
their confidence estimates remain imperfectly calibrated.
Our finding that low-temperature generation ($T{=}0.2$) does not yield
stable outputs complements this line of work: even when the sampling
distribution is sharply peaked, the resulting text varies substantially
across runs, suggesting that calibration and output determinism are
distinct challenges.

\paragraph{Medical LLM evaluation.}
Singhal et~al.~\cite{singhal2023large} demonstrated that large language
models encode clinical knowledge sufficient to approach expert-level
performance on medical licensing exams, while
Nori et~al.~\cite{nori2023gpt4med} showed that generalist foundation
models can match or exceed domain-specialized systems on medical
benchmarks.
However, both studies evaluate single-pass accuracy and do not report
run-to-run variance, leaving the reproducibility dimension of clinical
LLM behavior unexplored. Our work addresses this gap by treating
within-model consistency as a co-equal evaluation axis alongside
accuracy.

\paragraph{LLM-as-judge evaluation.}
Zheng et~al.~\cite{zheng2023judging} established the LLM-as-judge
paradigm for scalable open-ended evaluation, demonstrating strong
agreement with human raters while also documenting systematic biases
such as position and verbosity effects.
We adopt a locally-hosted open-source judge to maintain full
reproducibility and keep evaluation costs low, while acknowledging the
inherent stochasticity of the judge itself as a limitation
(see Discussion).

% ---- 3. Methodology -----------------------------------------
\section*{Materials and methods}

\subsection*{Datasets}

We use \textbf{MedQuAD}~\cite{abacha2019mqa}, a collection of
16,412 medical question-answer pairs curated from fourteen NIH online
health resources (NIHSeniorHealth, Genetics Home Reference, Cancer.gov,
and others).
Questions cover a wide range of clinical focus areas including symptoms,
treatments, diagnoses, prevention, and prognosis.
All content is in English and targeted at informed general audiences,
making it a good proxy for consumer-facing clinical QA.
Although MedQuAD is professionally curated rather than user-generated,
its questions closely mirror the types of health inquiries commonly
posted on platforms such as Reddit medical
communities~\cite{sager2021reddit}, covering symptoms, treatments,
and prevention in lay-accessible language. We therefore use MedQuAD as
a \emph{proxy benchmark} for the Reddit-style consumer health questions
that motivate this work: it provides clinically validated reference
answers (which user-generated forum posts lack) while remaining
representative of lay health information needs.

To test whether our findings generalize beyond a single corpus, we add
a second benchmark, \textbf{MedicationQA}~\cite{abacha2019medication}, a
set of 689 consumer medication questions (covering dosing, interactions,
administration, and side effects) paired with answers curated from
authoritative drug-information sources (the MedInfo~2019 collection).
MedicationQA is narrower and more drug-specific than MedQuAD and its
reference answers are short and factual, providing a complementary, less
forgiving stress test. We apply the identical sampling, inference, and
scoring pipeline to both datasets.

For reproducible sampling, we draw question subsets using a fixed
random seed (seed\,=\,42) so that any researcher can reconstruct the
exact evaluation set. To make this guarantee precise, we document the
full deterministic data-loading and sampling pipeline: we load a
version-pinned copy of the corpus (16{,}412 records; 14{,}984 distinct
question strings), apply no question removal or preprocessing, and draw
the subset with a single pandas \texttt{DataFrame.sample(random\_state=42)}
call over the loaded frame. Because the reproducibility of seeded
sampling depends on the dataset version, row ordering, and the specific
sampling implementation, we additionally provide the exact sampled
question set as Supporting Information (S1 File).
We evaluate $N{=}50$ randomly sampled questions \emph{per dataset}, each
subjected to 10 repeated inference runs per model, yielding 500
responses per model per dataset (2{,}500 per dataset; 5{,}000 in total
across the five models and two datasets).

\subsection*{Models and inference setup}

We select four locally-deployable open-weight models spanning
4B--12B parameters and two domain-specialization strategies, all served
via the Ollama runtime~\cite{ollama}, and include one closed-source API
model as an external reference point:

\begin{itemize}
  \item \textbf{Llama 3.1 8B}~\cite{dubey2024llama3}:
        Meta AI's instruction-tuned general-purpose model.
        Represents the dominant open-weight baseline at the 8B scale.
  \item \textbf{Gemma 3 12B}~\cite{team2024gemma}:
        Google DeepMind's instruction-tuned model at 12B parameters.
        Provides a larger general-purpose comparison point.
  \item \textbf{Gemma 3 4B}~\cite{team2024gemma}:
        The 4B instruction-tuned sibling of Gemma 3 12B. It serves a
        dual role: a \emph{scale} control against Gemma 3 12B and,
        crucially, a \emph{same-size} general-purpose control against
        MedGemma 1.5 4B, which together isolate model scale from
        clinical fine-tuning.
  \item \textbf{MedGemma 1.5 4B}~\cite{medgemma2024}:
        A Gemma-based 4B model fine-tuned on clinical corpora,
        testing the hypothesis that domain specialization compensates
        for smaller scale---now tested directly against its same-size
        base model, Gemma 3 4B.
  \item \textbf{GPT-4.1-nano} (closed-source reference):
        OpenAI's smallest GPT-4.1-tier model, accessed via the vendor
        API. We include it solely as an external reference to contextualize
        the open models, \emph{not} as a locally-deployable peer; it is
        therefore reported separately and excluded from ``best open
        model'' comparisons. Because closed APIs are versioned and may
        change without notice and their seed support is best-effort, we
        make no determinism claims for this model and run it only under
        the same decoding settings as the open models.
\end{itemize}

All models receive an identical, fixed system instruction:
\textit{``You are a clinical expert. Answer in $\leq$6\,sentences.
Be direct. If uncertain, say so. Never recommend unsafe actions.''}
Holding the system prompt constant isolates model behavior from
prompt-engineering effects, which is essential for a fair comparison.

Generation parameters are fixed across all models and all runs:
temperature $T{=}0.2$, top-$p{=}1.0$, max\,tokens\,$=512$,
request timeout\,=\,120\,s with two retries on failure.
The choice of $T{=}0.2$ is deliberate: this is a common production
default intended to reduce variance.
Our key research question is how much variance \emph{remains} even
in this near-deterministic regime.

\subsection*{Evaluation metrics}
\label{sec:metrics}

We organize metrics into three groups.

\paragraph{Quality metrics (vs.\ medically validated ground-truth).}
Each generated response is compared to the gold answer using six
complementary metrics:
(1)~\textbf{Exact Match}: binary equality after text normalization;
(2)~\textbf{Token F1}: token-level precision/recall F1 on
lowercased alphanumeric tokens;
(3)~\textbf{String Similarity}: character-level SequenceMatcher ratio;
(4)~\textbf{BLEU}~\cite{papineni2002bleu}: smoothed sentence-level
$n$-gram overlap;
(5)~\textbf{ROUGE-L}~\cite{lin2004rouge}: longest common subsequence F1;
(6)~\textbf{BERTScore F1}~\cite{zhang2020bertscore}: contextual
embedding similarity computed with \texttt{roberta-base}.
Exact match and token F1 reward lexical fidelity;
BERTScore captures semantic equivalence even when the phrasing differs.

\paragraph{LLM-as-judge score.}
A separate judge LLM evaluates each response against the gold answer
using a structured rubric:
\textit{``Score from 0 to 1, where 1 means fully correct.
Assess factual correctness and clinical safety.''}
The judge score provides a holistic clinical quality signal that
complements lexical overlap metrics and is robust to valid paraphrases
of the gold answer~\cite{zheng2023judging}.
Concretely, we use \texttt{gpt-oss:20b}, a locally-hosted 20B-parameter
open-source model served via Ollama, as the judge.
The judge inherits the same generation parameters as the target models
($T{=}0.2$, top-$p{=}1.0$, max\,tokens\,$=512$) and is invoked once
per response via the \texttt{/api/generate} endpoint (i.e., a single
judge pass with no averaging across multiple runs).
Scores are extracted from the judge output via regular-expression
parsing with a fallback heuristic.
Because the judge is itself stochastic, the per-response judge scores in
the main runs carry variance. To quantify the judge's own reliability,
we additionally re-judge a model-stratified subset of responses five
times and report the within-response score standard deviation and the
one-way intraclass correlation coefficient (ICC(1), the standard
test--retest reliability statistic) in the Discussion.

\paragraph{Reproducibility metrics.}
For each (model, question) pair, we compute the following metrics over
the $N$ repeated runs:

\begin{itemize}
  \item \textbf{Normalized self-agreement rate} (lexical): fraction of
        runs whose normalized output matches the modal (most frequent)
        normalized output.
        A value of 1.0 indicates perfect consistency; $1/N$ indicates
        every run produced a distinct response.
  \item \textbf{Normalized response uniqueness rate} (lexical): number
        of distinct normalized outputs divided by $N$.
        A value of 1.0 indicates every run was unique.
  \item \textbf{Semantic self-similarity}: the mean pairwise
        semantic similarity, measured as BERTScore-F1~\cite{zhang2020bertscore},
        across the $N$ run outputs for a given (model, question) pair.
        Unlike the two lexical measures, this credits runs that are
        differently worded but semantically equivalent, directly
        addressing the concern that paraphrase inflates apparent
        run-to-run variability. We caution that embedding-based
        similarity has a high baseline floor, so it should be read as a
        complement to---not a replacement for---the lexical measures.
\end{itemize}

The two lexical metrics operate on text normalized by lowercasing,
whitespace collapsing, and removal of non-alphanumeric characters,
making them robust to trivial formatting differences; the semantic
metric operates on the raw generated text.
Together these metrics are complementary: self-agreement emphasizes
the \emph{stability} of the most likely output, uniqueness
measures the \emph{breadth} of the output distribution, and semantic
self-similarity isolates \emph{meaning-level} agreement from surface
phrasing.

\paragraph{Efficiency.}
We record per-run latency (ms) and throughput (tokens/second)
from the inference server, as practical deployment constraints
frequently make efficiency a decisive factor.

\subsection*{Pipeline architecture}

The pipeline is implemented as a Python package with three stages:

\begin{itemize}
  \item Run: queries each model for each question $N$ times and stores
        raw responses as JSONL/CSV/Parquet.
  \item Score: applies the full metric suite to every stored response.
  \item Report: aggregates scored responses and renders the Markdown
        report.
\end{itemize}

The separation of generation from scoring is a key design choice:
it allows the scoring step (including new metrics) to be re-run
without re-querying the LLMs, which is expensive and introduces
additional variance into the experiment.
The pipeline is fully resumable: interrupted runs continue from
where they left off, making it practical on commodity hardware
where long jobs may be preempted.
All raw outputs are retained as an audit trail, enabling downstream
re-analysis without re-running models.
Inference is fully stateless across questions and runs: each query is
an independent, single-turn generation request with no conversation
history carried over between queries, so there is no cross-sample
context leakage that could otherwise inflate or deflate measured
reproducibility.

A shared system prompt, fixed random seed, and version-pinned
dependencies ensure that the exact experimental conditions can be
reproduced.

% ---- 4. Results ---------------------------------------------
\section*{Results}
\label{sec:results}

\begin{figure}[!ht]
\centering
\includegraphics[width=\textwidth]{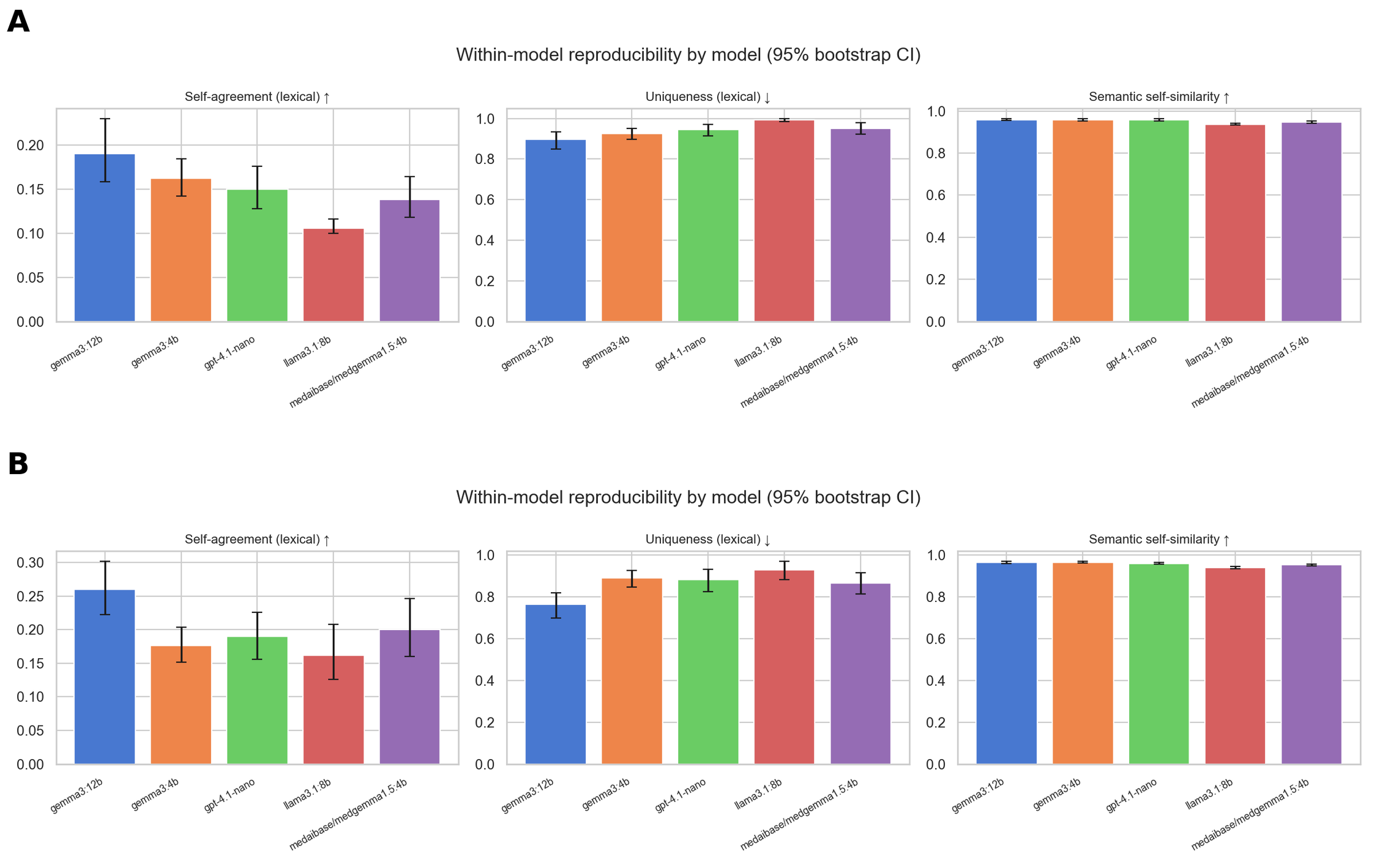}
\caption{{\bf Within-model reproducibility by model, with 95\% bootstrap
confidence intervals.} Per-model lexical self-agreement ($\uparrow$), lexical
uniqueness ($\downarrow$), and semantic self-similarity ($\uparrow$) across the
10 runs. \textbf{(A)} MedQuAD (top); \textbf{(B)} MedicationQA (bottom). Across
both datasets and all five models, lexical agreement is low while semantic
self-similarity is high (0.94--0.97), showing that run-to-run divergence is
largely paraphrastic rather than meaning-level.}
\label{fig:repro}
\end{figure}

\begin{figure}[!ht]
\centering
\includegraphics[width=\textwidth]{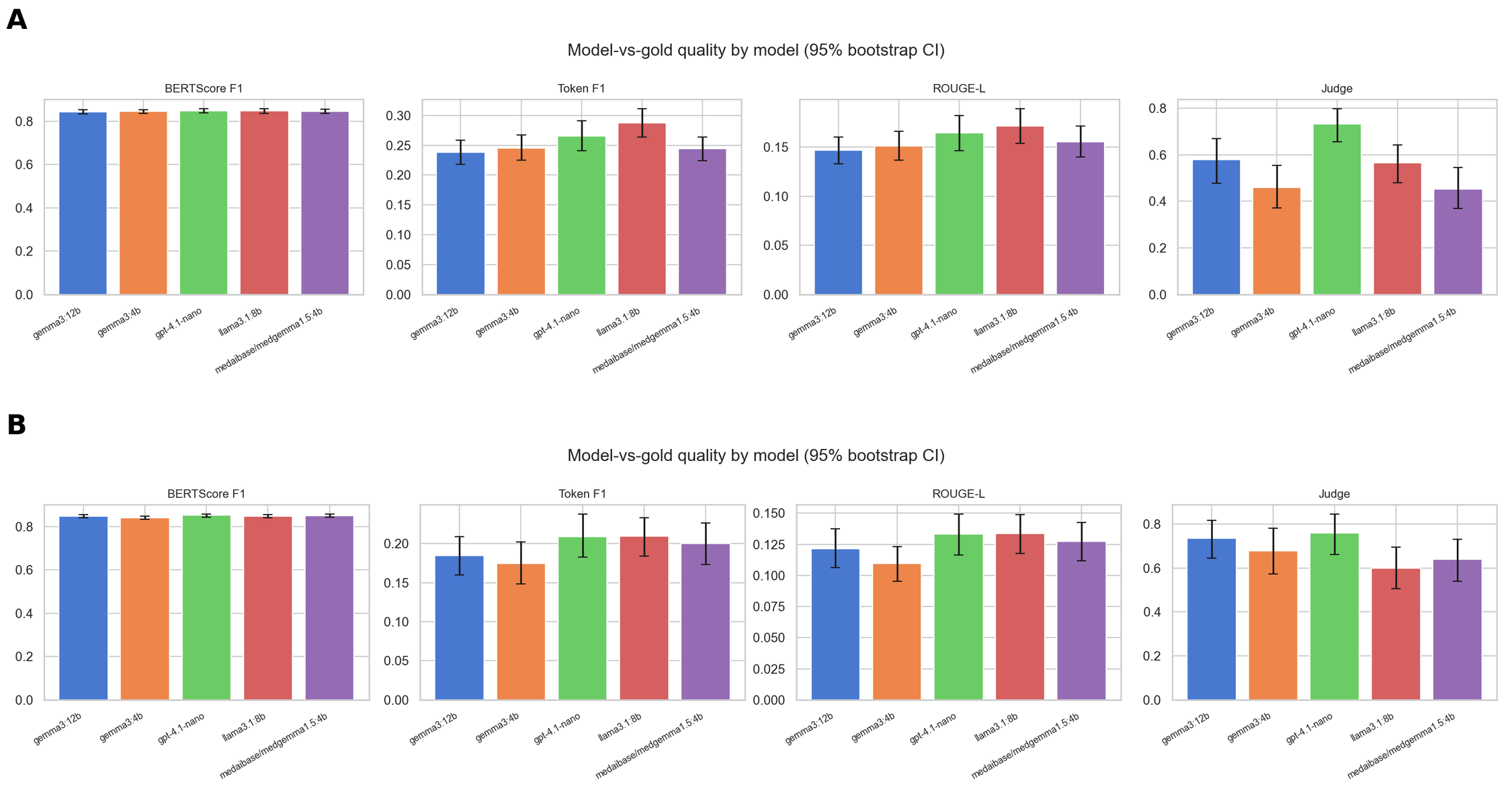}
\caption{{\bf Model-vs-gold quality by model, with 95\% bootstrap confidence
intervals.} BERTScore F1, Token F1, ROUGE-L, and LLM-judge score per model.
\textbf{(A)} MedQuAD (top); \textbf{(B)} MedicationQA (bottom). BERTScore is
tightly clustered across models; the closed reference GPT-4.1-nano leads on the
judge score but remains within the open models' range on the other metrics.}
\label{fig:quality}
\end{figure}

Tables~\ref{tab:quality} and~\ref{tab:repro} summarize results across
both datasets (50 questions $\times$ 10 runs $\times$ 5 models per
dataset; 2{,}500 responses per dataset, 5{,}000 in total), and
Figs~\ref{fig:repro} and~\ref{fig:quality} visualize the same
model-level estimates with their bootstrap confidence intervals.

\paragraph{Aggregation and uncertainty.}
We state the aggregation procedure explicitly. For the quality metrics,
a model's score is the mean over all of its responses; because the
design is balanced (10 runs per question), this equals the mean across
questions of the per-question mean. For the reproducibility metrics, we
compute each metric per (model, question) and then average over
questions. To avoid over-interpreting small between-model differences,
every model-level value is reported with a 95\% confidence interval
obtained by a clustered percentile bootstrap that resamples questions
with replacement (1{,}000 resamples); per-question standard deviations
are additionally available. For clarity, the bracketed quantities
reported throughout are 95\% bootstrap confidence intervals for the mean
(i.e., the lower and upper interval bounds), not observed
minimum--maximum ranges; per-question standard deviations for each
metric are provided in the supplementary material.

\begin{table}[!ht]
\begin{adjustwidth}{-0.75in}{-0.75in}
\centering
\caption{{\bf Quality metrics: model output vs.\ gold answer, mean [95\% CI].}
Per dataset, $n{=}50$ questions $\times$ 10 runs. Each value is the
across-question mean with a clustered percentile-bootstrap 95\% confidence
interval (1{,}000 resamples, resampling questions). Bracketed values are
95\% bootstrap CIs for the mean, not min--max ranges. Bold = best among the
open-weight models per column, per dataset.}
\label{tab:quality}
\footnotesize
\begin{tabular}{lcccc}
\toprule
\textbf{Model} & \textbf{BERTScore F1} & \textbf{Token F1} & \textbf{ROUGE-L} & \textbf{Judge} \\
\midrule
\multicolumn{5}{l}{\emph{MedQuAD}}\\
Llama 3.1 8B      & \textbf{0.848} [0.837, 0.857] & \textbf{0.287} [0.264, 0.311] & \textbf{0.172} [0.154, 0.189] & 0.565 [0.480, 0.643] \\
Gemma 3 12B       & 0.843 [0.833, 0.852] & 0.238 [0.218, 0.258] & 0.147 [0.133, 0.160] & \textbf{0.580} [0.477, 0.669] \\
Gemma 3 4B        & 0.845 [0.836, 0.854] & 0.245 [0.225, 0.267] & 0.151 [0.137, 0.166] & 0.460 [0.371, 0.554] \\
MedGemma 1.5 4B   & 0.846 [0.835, 0.855] & 0.244 [0.224, 0.263] & 0.155 [0.140, 0.172] & 0.452 [0.369, 0.545] \\
GPT-4.1-nano$^\dagger$ & 0.848 [0.838, 0.859] & 0.266 [0.241, 0.291] & 0.165 [0.146, 0.182] & 0.733 [0.656, 0.798] \\
\midrule
\multicolumn{5}{l}{\emph{MedicationQA}}\\
Llama 3.1 8B      & 0.849 [0.841, 0.855] & \textbf{0.209} [0.184, 0.233] & \textbf{0.134} [0.118, 0.149] & 0.600 [0.507, 0.696] \\
Gemma 3 12B       & 0.848 [0.841, 0.855] & 0.185 [0.160, 0.209] & 0.121 [0.106, 0.137] & \textbf{0.735} [0.646, 0.816] \\
Gemma 3 4B        & 0.842 [0.834, 0.848] & 0.175 [0.148, 0.202] & 0.110 [0.095, 0.123] & 0.680 [0.573, 0.781] \\
MedGemma 1.5 4B   & \textbf{0.851} [0.843, 0.858] & 0.200 [0.174, 0.226] & 0.127 [0.112, 0.143] & 0.640 [0.539, 0.731] \\
GPT-4.1-nano$^\dagger$ & 0.852 [0.844, 0.859] & 0.209 [0.182, 0.238] & 0.133 [0.116, 0.150] & 0.760 [0.662, 0.847] \\
\bottomrule
\end{tabular}
\\[2pt]
{\footnotesize $^\dagger$ Closed-source API model (OpenAI), included as an
external reference point, not as a peer of the locally-deployable open-weight
models; excluded from the per-column ``best'' comparison.}
\end{adjustwidth}
\end{table}

\begin{table}[!ht]
\begin{adjustwidth}{-0.75in}{-0.75in}
\centering
\caption{{\bf Within-model reproducibility across 10 runs, mean [95\% CI].}
Self-agreement (higher is better; $1/N{=}0.10$ = every run distinct) and
uniqueness (lower is better) are lexical; semantic self-similarity is the mean
pairwise BERTScore-F1 across the 10 runs (higher is better, robust to
paraphrase). Bracketed values are 95\% bootstrap CIs for the mean, not
min--max ranges. Bold = best among the open-weight models per column, per
dataset.}
\label{tab:repro}
\footnotesize
\begin{tabular}{lccc}
\toprule
\textbf{Model} & \textbf{Self-agreement} $\uparrow$ & \textbf{Uniqueness} $\downarrow$ & \textbf{Semantic self-sim.} $\uparrow$ \\
\midrule
\multicolumn{4}{l}{\emph{MedQuAD}}\\
Llama 3.1 8B      & 0.106 [0.100, 0.116] & 0.994 [0.984, 1.000] & 0.937 [0.932, 0.941] \\
Gemma 3 12B       & \textbf{0.190} [0.158, 0.230] & \textbf{0.896} [0.848, 0.934] & \textbf{0.959} [0.954, 0.964] \\
Gemma 3 4B        & 0.162 [0.142, 0.184] & 0.926 [0.898, 0.950] & 0.958 [0.953, 0.962] \\
MedGemma 1.5 4B   & 0.138 [0.118, 0.164] & 0.952 [0.922, 0.978] & 0.947 [0.942, 0.952] \\
GPT-4.1-nano$^\dagger$ & 0.150 [0.128, 0.176] & 0.944 [0.914, 0.970] & 0.958 [0.953, 0.962] \\
\midrule
\multicolumn{4}{l}{\emph{MedicationQA}}\\
Llama 3.1 8B      & 0.162 [0.126, 0.208] & 0.930 [0.882, 0.972] & 0.939 [0.934, 0.944] \\
Gemma 3 12B       & \textbf{0.260} [0.222, 0.302] & \textbf{0.764} [0.700, 0.820] & \textbf{0.965} [0.959, 0.970] \\
Gemma 3 4B        & 0.176 [0.152, 0.204] & 0.892 [0.848, 0.928] & \textbf{0.965} [0.960, 0.969] \\
MedGemma 1.5 4B   & 0.200 [0.160, 0.246] & 0.868 [0.814, 0.916] & 0.952 [0.946, 0.957] \\
GPT-4.1-nano$^\dagger$ & 0.190 [0.156, 0.226] & 0.884 [0.826, 0.932] & 0.959 [0.955, 0.964] \\
\bottomrule
\end{tabular}
\\[2pt]
{\footnotesize $^\dagger$ Closed-source API reference (see
Table~\ref{tab:quality}); excluded from the per-column ``best'' comparison.}
\end{adjustwidth}
\end{table}

\begin{table}[!ht]
\centering
\caption{{\bf Inference efficiency (MedQuAD, tokens/second).} Median per-run
throughput; MedicationQA throughput is comparable. GPT-4.1-nano is API-served
and network-bound rather than compute-bound on local hardware.}
\label{tab:efficiency}
\begin{tabular}{lc}
\toprule
\textbf{Model} & \textbf{Tokens/s} \\
\midrule
GPT-4.1-nano$^\dagger$ & 97.9 \\
Gemma 3 4B        & 61.0 \\
Llama 3.1 8B      & 41.3 \\
MedGemma 1.5 4B   & 29.1 \\
Gemma 3 12B       & 23.7 \\
\bottomrule
\end{tabular}
\end{table}

\paragraph{Quality.}
Across both datasets, BERTScore F1 is tightly clustered (0.84--0.85)
for all five models, indicating broadly similar embedding-level
alignment with the gold answers and limited discriminative power at the
semantic-similarity level. The lexical and judge metrics separate the
models more clearly. On MedQuAD, Llama 3.1 8B leads the open models on
Token F1 (0.287) and ROUGE-L (0.172), whereas Gemma 3 12B obtains the
highest open-model judge score (0.580); on MedicationQA the judge scores
are higher overall and Gemma 3 12B again leads the open models (0.735).
The closed-source reference, GPT-4.1-nano, obtains the highest judge
score on both datasets (0.733 and 0.760), but its BERTScore, Token F1,
and ROUGE-L all fall within the range spanned by the open models, so it
does not dominate the comparison---a useful calibration point rather
than an upper bound the open models fail to approach.

A closer look at the open-versus-closed comparison clarifies why the
closed reference does not dominate. Token F1 and ROUGE-L measure surface
overlap---shared $n$-grams and longest common subsequence---between a
response and the gold answer, so they reward outputs phrased close to the
reference wording rather than outputs that are merely correct. On
MedQuAD, whose gold answers are drawn from NIH consumer-health pages,
Llama 3.1 8B in fact exceeds GPT-4.1-nano on both Token F1 (0.287 vs.\
0.266) and ROUGE-L (0.172 vs.\ 0.165), yet GPT-4.1-nano obtains the
higher judge score (0.733 vs.\ 0.565). The two signals are not in
conflict: GPT-4.1-nano tends to reformulate and restructure its answers,
increasing lexical distance from the reference string and depressing
$n$-gram overlap, while the judge---scoring factual correctness and
clinical safety rather than wording---still rates its content highest.
BERTScore is tightly clustered (0.84--0.85), so embedding-level alignment
is comparable across models. The separation therefore occurs on two
distinct axes: a surface-lexical axis, on which the closest paraphraser
wins, and a holistic-quality axis, on which the closed reference leads.
Accordingly, we do not read Token F1/ROUGE-L as quality rankings; we
interpret them as proximity-to-reference and read model quality
primarily from the judge score.

Notably, no model produces an exact lexical match with any gold
answer (exact match = 0.000 across all models), confirming that all
models paraphrase rather than recall; we also read this as a signal of
no overfitting to the medically validated ground-truth corpus.

\paragraph{Reproducibility.}
The central finding is a large, consistent gap between \emph{lexical}
and \emph{semantic} reproducibility that holds across both datasets and
all five models, including the closed reference (Fig~\ref{fig:repro}).
Lexically, outputs are highly variable: self-agreement ranges from 0.11
to 0.26 and uniqueness from 0.76 to 0.99, so even the most consistent
model reproduces its modal answer on only a minority of runs (e.g., on
MedQuAD, Llama 3.1 8B is unique on 99.4\% of runs and Gemma 3 12B on
89.6\%). Semantically, however, the same runs are highly similar: mean
pairwise semantic self-similarity is 0.94--0.97 for every model on both
datasets. In other words, run-to-run divergence at $T{=}0.2$ is
overwhelmingly \emph{paraphrastic} rather than meaning-level---models
say substantially the same thing in different words. This directly
addresses the concern that purely lexical metrics overstate instability
by counting paraphrases as disagreements; we nonetheless read the
semantic figure as a complement to, not a replacement for, the lexical
measures, given the known high baseline floor of embedding similarity.
Pairwise \emph{exact} output overlap across model pairs is 0.000, which
we treat only as a qualitative observation: strict string comparison is
uninformative about semantic or clinical equivalence---two medically
equivalent but differently worded answers count as non-overlapping. We
therefore frame these results as run-to-run \emph{output variability},
not as direct evidence of clinical inconsistency.

\paragraph{Quality--reproducibility trade-off and the role of scale.}
A trade-off is again visible among the open models. Llama 3.1 8B attains
the highest lexical quality and throughput on MedQuAD but is the least
lexically reproducible (uniqueness 0.99) and has the lowest semantic
self-similarity. Gemma 3 12B is the most reproducible model on both
datasets---on both the lexical and the semantic measures---while also
achieving the best open-model judge score, at the cost of the lowest
throughput (Table~\ref{tab:efficiency}). With the same-size control now
included, the effect of scale is clean: Gemma 3 12B is more reproducible
and higher-scoring than Gemma 3 4B. The clinically fine-tuned
MedGemma 1.5 4B, by contrast, is not statistically distinguishable from its
same-size general-purpose sibling Gemma 3 4B: it matches or trails on the
judge score (0.452 vs.\ 0.460 on MedQuAD; 0.640 vs.\ 0.680 on MedicationQA)
and, although its point estimates edge ahead on some lexical and BERTScore
measures on MedicationQA, all of these differences fall within overlapping
confidence intervals. At these scales it is therefore capacity, not domain
fine-tuning, that drives the distinguishable differences (see Discussion).

\subsection*{Qualitative case studies}

Beyond the aggregate metrics, we examine representative individual items
to make the numeric patterns concrete and to address what practitioners
can learn from specific successes and failures. Because the LLM judge is
not a clinical ground truth---its test--retest reliability is only
moderate on MedicationQA (ICC(1)~$=$~0.69; see Discussion)---we frame
these cases as differences in judge score and in alignment with the gold
reference, not as independent clinical adjudications. Full verbatim
outputs for all 10 runs and the complete ranked candidate lists
underlying these examples are provided as Supporting Information
(S2 File).

\paragraph{Higher lexical overlap did not mean a better answer.}
The opposite pattern---surface-overlap metrics favoring an answer the
judge rated poorly---appears on the MedQuAD item on the genetic causes of
hereditary multiple osteochondromas. MedGemma 1.5 4B scored higher than
GPT-4.1-nano on both Token F1 (0.27 vs.\ 0.25) and ROUGE-L (0.16 vs.\
0.15), yet received a mean judge score of 0.00 against GPT-4.1-nano's
0.92. The open model attributed the condition to the \textit{GNAS} gene,
whereas GPT-4.1-nano named \textit{EXT1} and \textit{EXT2}---the genes
given in the gold answer. Here the higher $n$-gram overlap reflected
shared generic phrasing rather than a correct answer, a concrete instance
of the surface-versus-quality separation quantified above and a caution
against reading Token F1 or ROUGE-L as quality rankings.

\paragraph{Run-to-run variation is paraphrastic, not contradictory.}
To make the lexical--semantic gap concrete, consider Gemma 3 4B on the
MedQuAD item ``What is (are) Afibrinogenemia?''. Across all 10 runs every
response is lexically unique (uniqueness 1.00, self-agreement 0.10), yet
their mean pairwise semantic self-similarity is 0.98 and the mean judge
score is 0.95. Each run states near-identical content in shifting
words---e.g., ``\emph{\ldots an extremely rare inherited bleeding
disorder characterized by a complete or near-complete absence of
fibrinogen\ldots}'' with one run continuing ``\emph{\ldots often
manifesting in infancy}'' and another ``\emph{\ldots often into joints
and muscles}''---so the clinical substance is preserved while the
surface form changes. This is the item-level signature of the aggregate
pattern in Table~\ref{tab:repro}: low lexical agreement co-occurring with
high semantic stability.

\paragraph{Reference-based judging can reward the less cautious answer.}
A final case illustrates why we treat the judge as a useful but fallible
signal rather than a ground truth. On the MedicationQA item ``Can
alprazolam lower blood pressure?'', Gemma 3 4B received a mean judge
score of 0.83 versus 0.04 for GPT-4.1-nano---the largest such gap on
MedicationQA. Gemma 3 4B answered affirmatively (``\emph{Yes, alprazolam,
a benzodiazepine, can lower blood pressure\ldots}''), matching the
direction of the gold reference (``\emph{Alprazolam is as effective as
captopril in lowering BP in ED patients with an initial SBP $>$
160\,mmHg}''), whereas GPT-4.1-nano gave the more cautious answer that
alprazolam ``\emph{is not typically used or recommended as a treatment
for hypertension}''. Because the judge scores agreement with the gold
answer, it rewarded the response aligned with a single, narrowly scoped
reference over the more conservative one---an outcome reference-based
grading can produce whenever the reference itself is narrow. We
therefore present this not as evidence that the open model is clinically
superior, but as a concrete instance of the judge-reliability limitation
we quantify in the Discussion (test--retest ICC(1)~$=$~0.69 on
MedicationQA); it reinforces our recommendation that LLM outputs in
clinical settings receive human-in-the-loop review rather than be
accepted on a judge score alone.

Together these cases show that model rankings depend on which axis is
measured---lexical proximity to the reference, holistic judged quality,
or run-to-run stability---and that no single metric captures
deployment-relevant quality on its own.

% ---- 5. Discussion ------------------------------------------
\section*{Discussion}
\label{sec:discussion}

\paragraph{Reproducibility is not solved by low temperature.}
The most actionable finding is that $T{=}0.2$ does not yield
\emph{lexically} reproducible behavior in practice.
Self-agreement rates of 0.11--0.26 mean that for a given clinical
question, the same model almost always returns differently worded text
on the next invocation: Llama 3.1 8B, the least consistent model,
produces a unique normalized response on up to 99\% of runs.
Importantly, this instability is largely surface-level: semantic
self-similarity across runs stays high (0.94--0.97), so the dominant
risk is variable phrasing and emphasis rather than outright
contradictory clinical content. Even so, variable phrasing can shift
perceived meaning, complicate auditing, and undermine user trust, so for
clinical deployment we still recommend output-stabilization
strategies: majority voting across multiple runs, confidence-score
gating, or mandatory human-in-the-loop review before any LLM output
is surfaced to a clinician.
Single-pass benchmarks that report average accuracy without also
reporting within-model variance conceal this risk entirely. Here
we would also like to emphasize that we did not use $T{=}0.0$ not
only because it is not common for question answering GenAI tasks,
but also as a gatekeeper of training bias in case the dataset
of medically validated ground-truth or portions of it was visible
during the model(s) training.

\paragraph{Clinical fine-tuning versus model scale.}
Including Gemma 3 4B as a same-size general-purpose control lets us
separate domain fine-tuning from model scale---the comparison the
original study could only flag as a confound. Two patterns emerge.
First, \emph{scale helps}: Gemma 3 12B outperforms Gemma 3 4B on judge
score and is more reproducible on both datasets. Second, at fixed size,
\emph{clinical fine-tuning is not distinguishable from its base}: MedGemma
1.5 4B matches or trails its same-size base model Gemma 3 4B on the judge
score (0.452 vs.\ 0.460 on MedQuAD; 0.640 vs.\ 0.680 on MedicationQA), and
although its point estimates exceed Gemma 3 4B on several lexical and
reproducibility measures on MedicationQA, every such difference lies within
overlapping confidence intervals. The earlier apparent ``MedGemma
underperforms'' effect was therefore largely a capacity effect: once size is
held constant, domain adaptation confers no statistically distinguishable
benefit on these consumer-health QA tasks.
This is consistent with the broader observation that scale often
dominates domain adaptation when the target task is well covered in
pretraining~\cite{singhal2023large, nori2023gpt4med}.
Practitioners should evaluate models empirically rather than relying
on domain-specific branding, and should prefer same-scale comparisons
to isolate the contribution of fine-tuning.

\paragraph{LLM-as-judge introduces its own variance.}
The judge model (a 20B open-source model, $T{=}0.2$) provides a
clinically meaningful holistic quality signal, but is itself
stochastic, and in the main runs each response was judged once.
To characterize this, we re-judged a model-stratified subset of
responses five times on each dataset. The judge was highly stable on
MedQuAD---mean within-response score standard deviation 0.064 and
one-way intraclass correlation ICC(1)~$=$~0.94 ($n{=}22$
responses)---and moderately stable on MedicationQA (within-response SD
0.111, ICC(1)~$=$~0.69, $n{=}25$), indicating that most of the score
variance on MedQuAD, and a substantial part on the harder MedicationQA
items, reflects genuine between-response differences rather than judge
noise. We separately note that the judge returned an unparseable score
on a small fraction of responses (5.2\% on MedQuAD, 2.9\% on
MedicationQA); these were excluded from the judge means, and all other
metrics are unaffected. Because the judge score is also the metric on
which models diverge most, this residual variance remains a meaningful
limitation. For production evaluations, we recommend running the judge
multiple times per response and averaging, or adopting a deterministic
rubric-grading approach where feasible.

\subsection*{Toward improved reproducibility: type-3 ANFIS and related methods}
\label{sec:improve}

Our metrics \emph{diagnose} reproducibility; a natural next question is
how to \emph{improve} it. Several complementary directions exist, and we
position our contribution relative to them. \textbf{Self-consistency}
decoding~\cite{wang2022selfconsistency, chen2023universal} samples
multiple generations and returns the most frequent answer; it is the
output-stabilization analogue of the variability we measure, and our
repeated-inference protocol provides exactly the samples such a scheme
would aggregate. \textbf{Semantic entropy}~\cite{farquhar2024semantic}
clusters generations by meaning and estimates uncertainty in semantic
space to flag confabulations; our semantic self-similarity metric is a
lightweight, deployment-oriented cousin of this idea, trading a full
entropy estimate for an interpretable pairwise-agreement score.
\textbf{Conformal prediction}~\cite{angelopoulos2021conformal} offers
distribution-free guarantees and could wrap a model's repeated outputs
in calibrated prediction sets, converting raw variability into a
formal coverage statement suitable for safety-critical review.

A further promising direction, and one we highlight at the editor's
suggestion, is to model the higher-order uncertainty we observe with a
\textbf{type-3 fuzzy logic / ANFIS} layer. Type-3 fuzzy
systems extend type-1 and interval type-2 systems with an additional
level of membership-function uncertainty, which makes them well suited
to representing the ``uncertainty about uncertainty'' that arises when a
model returns many semantically related but non-identical answers across
runs~\cite{mohammadzadeh2020type3, cao2021type3}. Concretely, an
adaptive neuro-fuzzy inference system (ANFIS) with type-3 antecedents
could act as an uncertainty-aware aggregator over the $N$ repeated
generations, mapping features of the run set (e.g., pairwise semantic
similarity, judge-score spread, response-length dispersion) to a
calibrated stability/confidence score that gates whether an answer is
surfaced, escalated, or withheld. We leave a full implementation and
evaluation of such a type-3 ANFIS post-processing layer to future work,
but note it as a principled route from \emph{measuring} reproducibility
to \emph{actively managing} it in deployment.

\paragraph{Inference environment.}
All results were obtained via the Ollama local inference server on a
single workstation.
Quantization level, hardware configuration, and batch size can affect
generation quality and throughput; API-hosted counterparts of the same
model families may exhibit different variance profiles.

% ---- 6. Conclusion ------------------------------------------
\section*{Conclusion}

We have presented and applied a reproducibility-first evaluation
framework for small open-weight LLMs in clinical question answering,
covering 5,000 responses from five models (four open-weight, plus a
closed-source reference) evaluated on two consumer-health datasets
(MedQuAD and MedicationQA), 50 questions each with 10 runs.
The results lead to three concrete takeaways for practitioners.

First, \textbf{low temperature does not buy lexical reproducibility, but
the variability is mostly paraphrastic}.
Even at $T{=}0.2$, lexical self-agreement is only 0.11--0.26 and
76--99\% of outputs per model are unique; yet semantic self-similarity
stays at 0.94--0.97, so models largely restate the same content in
different words. Any medically related usage must still account for this
surface variance through ensembling, gating, or human review.

Second, \textbf{at these parameter counts, model scale---not clinical
fine-tuning---drives quality}.
With a same-size control (Gemma 3 4B) now included, MedGemma 1.5 4B
is statistically indistinguishable from its general-purpose sibling on both
datasets---every difference falls within overlapping confidence
intervals---while the larger Gemma 3 12B leads the open models. The earlier
apparent shortfall of the clinically fine-tuned model was largely a capacity
effect. Model selection should be empirically driven, not label-driven.

Third, \textbf{quality and reproducibility partially trade off}.
Llama 3.1 8B maximizes lexical quality and speed but minimizes
consistency; Gemma 3 12B offers the best balance of quality and
reproducibility at the cost of throughput.
The right choice depends on whether the deployment context prioritizes
answer fidelity, consistency, or latency.

Our pipeline makes this three-dimensional evaluation
accessible to any team with a commodity workstation, and we invite
the clinical NLP community to adopt reproducibility metrics as a
standard alongside accuracy in LLM benchmarking.

\paragraph{Future work.}
Planned extensions include additional models (Phi-3, Mistral-Medical,
Qwen-Med), evaluation on further benchmarks (e.g., USMLE-style MedQA),
ensemble and type-3 ANFIS decoding/aggregation strategies to actively
improve reproducibility, a human spot-check of the LLM judge against
expert clinician ratings, and evaluation on a real-world EHR-derived
question set (e.g., MIMIC-IV~\cite{johnson2023mimic}).

% ---- Supporting information ---------------------------------
\section*{Supporting information}

\paragraph*{S1 File.}
\label{S1_File}
{\bf Sampled evaluation question sets.} The exact 50 questions used for
evaluation from each dataset (MedQuAD and MedicationQA).

\paragraph*{S2 File.}
\label{S2_File}
{\bf Qualitative case studies: verbatim model outputs and ranked candidate
lists.} Supporting detail for the representative items discussed in the
Qualitative case studies section---the full question text, gold answers,
verbatim model responses across all 10 runs, and the associated per-item
judge, lexical, and semantic scores, together with the deterministic
ranked candidate lists from which the highlighted cases were selected.

% ---- References ---------------------------------------------

\end{document}